\documentclass[aps,rsi,graphicx,floatfix, preprint]{revtex4-2}

\usepackage{graphicx}
\usepackage{comment}
\usepackage[dvipsnames]{xcolor}
\usepackage{xcolor}

\usepackage{hyperref}

\usepackage{lineno}

\begin{document}


\title{
Effects of the vibrations and electromagnetic pollution induced by a closed-cycle pulse-tube cryostat on high-accuracy electrical measurements.} 



\author{M. Taupin}
\email[]{mathieu.taupin@lne.fr}
\author{K. Dougdag}
\author{D. Ziane}
\author{F. Cou\"edo}
\affiliation{LNE, 29 avenue Roger Hennequin, 78190 Trappes-France}


\date{\today}

\begin{abstract}
\textcolor{black}{A paradigm shift in achieving (sub-)Kelvin environments is currently underway with the democratization of cryogen-free cryocoolers. This transition is driven by their user-friendly, continuous operation and the elimination of liquid helium, which is becoming increasingly scarce and expensive. Thanks to their large sample space and high cooling power, these systems can host superconducting magnets, making them an ideal platform for quantum technologies, materials science, low-temperature detectors, and even medical applications.}
The drawback is that this type of systems is inherently based on gas compression that induces a certain level of vibrations and electromagnetic perturbations, which can potentially prevent the determination of low amplitude signals or spoil their stability. In this paper we demonstrate that pulse-tube based cryocoolers can be used for electrical precision measurements under high magnetic field, using a commercial cryomagnetic system combined with a customized coaxial cryoprobe. Parts-per-billion level of measurement uncertainties in resistance determination, based on the quantum Hall resistance standard, is achievable, \textcolor{black}{matching} the state-of-the-art involving conventional cryostats based on liquid helium. We performed an extensive characterization of the cryomagnetic system to determine the level of vibrations and electromagnetic perturbations, revealing 
\textcolor{black}{the drastic effect of the magnetic field on the electromagnetic noise level.} If interferences can happen in some measurements, others appear immune to the perturbations, meaning that the cryogen-free system is not inherently a source of disturbance. The set of characterization measurements presented here is easily implementable in laboratories, which can help to determine the vibrations and electromagnetic pollution generated by any cryocooler.
\end{abstract}

\maketitle 


\section{Introduction}
Cryogen-free cooling systems are based on compression-expansion gas thermodynamic cycles, which \textcolor{black}{allow to} reach temperatures as low as few kelvins without the need of liquid helium \cite{Radebaugh_2003}. Pushed by the continuous increase of the demand and cost of liquid helium, together with easy-to-use aspects without requiring particular facilities, these cryogenic systems are extremely popular in numerous branches, such as spectroscopic measurements \cite{Scholtz_2016, Barber_2024, Esser_2024}, medical field \cite{Ackermann_2002}, material and quantum science \cite{Kalra_2016}, space instruments \cite{Chen_2024}, particle detectors \cite{Lee_2018, DAddabbo_2018}, microkelvin experiments \cite{Batey_2013}, and metrological measurements \cite{Janssen_2015, Rigosi_2019, Chae_2024, Chae_2025}. 
\textcolor{black}{However, as thermal engines, these systems rely on compressors and inherently contain moving parts close to the sample space. Consequently, their mechanical vibrations and electromagnetic noise levels are significantly higher than those of conventional, liquid-helium-cooled cryostats} \cite{Tomaru_2004, Schmidt_2020, Sato_2025}. Without precautions, effects of these pollutions can be clearly observed on the measurements, and therefore technical solutions to reduce their influence are being developed, such as active and passive damping \cite{Lee_2018, Schmoranzer_2019, Guan_2024, Zhang_2025}, helium battery \cite{Franklin_2023}, active noise cancellation \cite{DAddabbo_2018}, close-cycle helium circulation \cite{Adachi_2016, Adachi_2017}, electromagnetic shielding and filtering \cite{Adachi_2016, Adachi_2017}, and post-processing signal analysis \cite{Oyama_2022}. Each of these solutions is specific to the measurement of interest, and to be deployed, a preliminary characterization of the pollution generated by the cryogen-free cryocooler is necessary.

Reducing these perturbations is \textcolor{black}{essential} for fundamental quantum metrology \textcolor{black}{which requires a cryogenic environment to operate the quantum electrical standards}. Since the revision of the International System of Units (SI) in 2019, the units are \textcolor{black}{defined} from fundamental constants and not anymore \textcolor{black}{from} specific practical realizations. The ohm \textcolor{black}{($\Omega$), unit of electric resistance, is thus defined} in terms of $h/e^2$, with $h$ the Planck constant and $e$ the elementary charge. Consequently, it is possible to \textcolor{black}{realize} the ohm directly using the quantum Hall effect (QHE) and the associated von Klitzing constant $R_{{K}}=h/e^2$, \textcolor{black}{for which low temperature and magnetic field are needed. It is therefore essential to characterize the impact of induced environmental perturbations (vibrations and electromagnetic pollutions) generated by the cryocooler to reduce them to a minimum.} 

In this work we demonstrate that cryogen-free systems are compatible with low-noise high-precision electrical measurements, and competitive with state-of-the-art liquid-helium-based cryostats. We provide an extensive study of the vibrations and electromagnetic perturbations, and their potential influence on the measurements. It appears that \textcolor{black}{the electromagnetic pollution from the cryocooler is strongly enhanced by the applied magnetic field and is present also at high frequency (hundreds of kilohertz). If these additional pollutions affect only marginally the accuracy and precision of our resistance measurements in the quasi-direct current (DC) regime, instabilities are observed when using our Superconducting QUantum Interference Device (SQUID)-based detector, in which the working frequency closely matches with these perturbation frequencies.} It is therefore essential to \textcolor{black}{investigate} the characteristic frequencies of both the perturbations and the instrumentation in order to take the adequate filtering and/or damping measures to make the most of cryogen-free systems. 

\section{Pulse-tube cryostat and facility overview}
\subsection{The cryomagnetic system}

The cryogen-free cryomagnetic system is a commercial \textcolor{black}{instrument} from Cryogenic Ltd., incorporating a pulse-tube cryocooler from Sumitomo (model SRP-082, associated with the cold head model RP-082B2S and the compressor model F-70). This cryocooler has the valve unit separated from the cold head (hereafter named ``remote motor''), adapted for low-vibration low-noise operation. The system is composed of an integrated variable temperature insert (VTI), compatible with a top-loading insert, with a closed-cycle (gaseous) helium circulation for an accessible temperature ranging from 1.25~K to 300~K together with a 14~T superconducting magnet.

\begin{figure*}[ht!]
    \centering
    \includegraphics[width=1\linewidth]{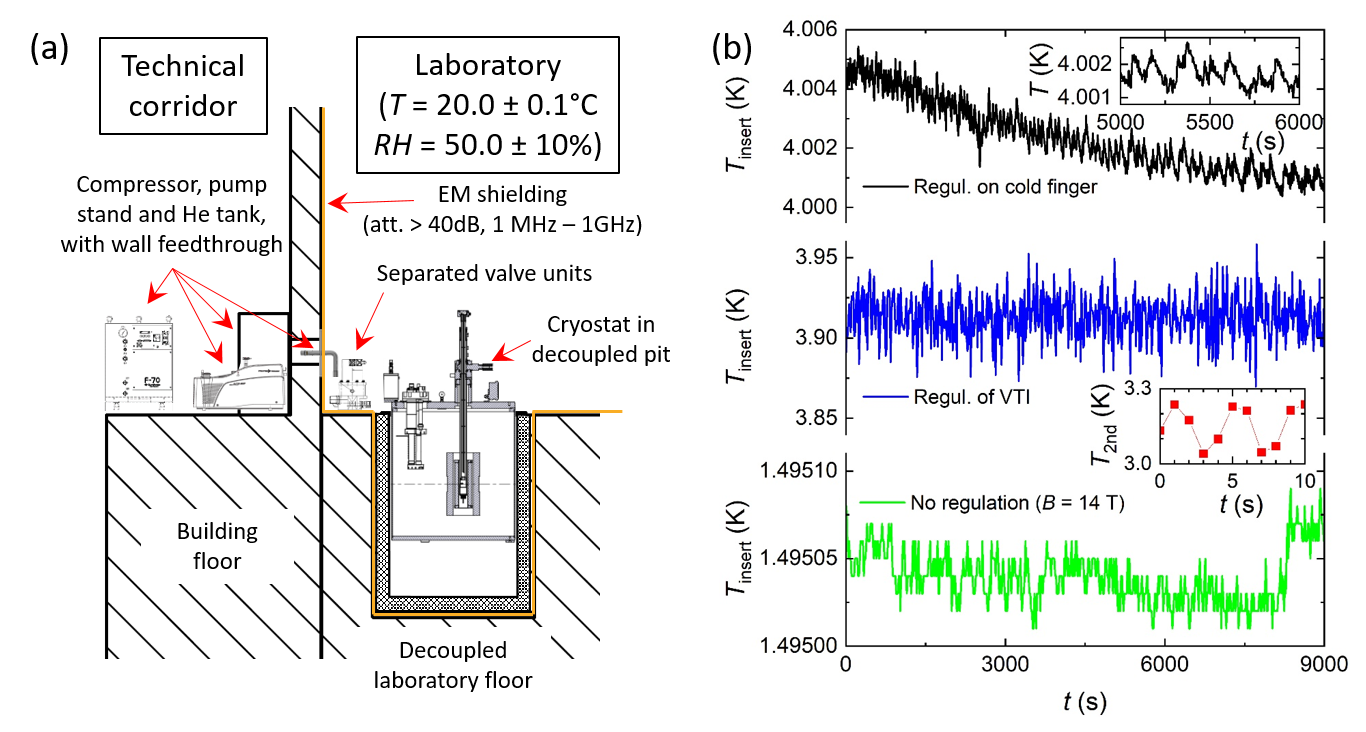}
    \caption{(a) Schematic view of the facility layout: On the right hand side is the laboratory, regulated in temperature and relative humidity (RH), with an electromagnetic shield (in yellow). The cryostat, in its own pit, has a separated valve unit. It is connected to the helium tank, pump station and compressor, situated in the technical corridor (left hand side) via wall feedthroughs. (b) Evolution of the temperature of the cold finger under various conditions: On top in black when regulated at 4~K using the heater of the cold finger, on a shorter timescale in insert; In the middle in blue when regulated at 4~K using the VTI heater; At the bottom in green at base temperature at 14~T (without heating). The typical evolution of the temperature of the second stage of the cold head is shown in the lower insert (red curve).}
    \label{Fig1}
\end{figure*}

The cryostat and the valve unit are situated in \textcolor{black}{a laboratory which offers excellent environmental conditions: electromagnetic shielding, regulated in temperature and relative humidity, several levels of mechanical decoupling, and low-pass filters on the electrical network}. The compressor and the pump stand are located in the technical corridor to limit the transmitted vibrations, and connected to the cryostat through wall feedthroughs (on a concrete wall) to guarantee the continuity of the electromagnetic shielding. A schematic view of the laboratory layout is shown in Fig.~\ref{Fig1}(a).

The cryostat presents excellent thermal performances and temperature stability, as presented in Fig.~\ref{Fig1}(b) showing the temperature of the insert (using a Cernox thermometer placed on the brass plate on the cold finger using the homemade insert presented in the next section). At the base temperature (green curve), the temperature dispersion is smaller than 0.1~mK, and when regulated at 4~K using a heater on the cold finger (black curve), it reaches 1~mK on a 1000~s time scale. For more convenience, the temperature is usually regulated using a heater at the bottom of the VTI column, at the expense of a larger temperature dispersion, of around 40~mK (blue curve). This \textcolor{black}{should} be compared to the oscillations of the temperature of the second stage of the cold head with an amplitude of approximately 0.25~K at 3~K (lower insert of Figure~\ref{Fig1}(b)). In the following, the temperature of the samples and the cold finger are assumed to be identical since they are always either under exchange gas or in liquid helium.

\subsection{Estimation of the mechanical vibrations}

\begin{figure*}
    \centering
    \includegraphics[width=1\linewidth]{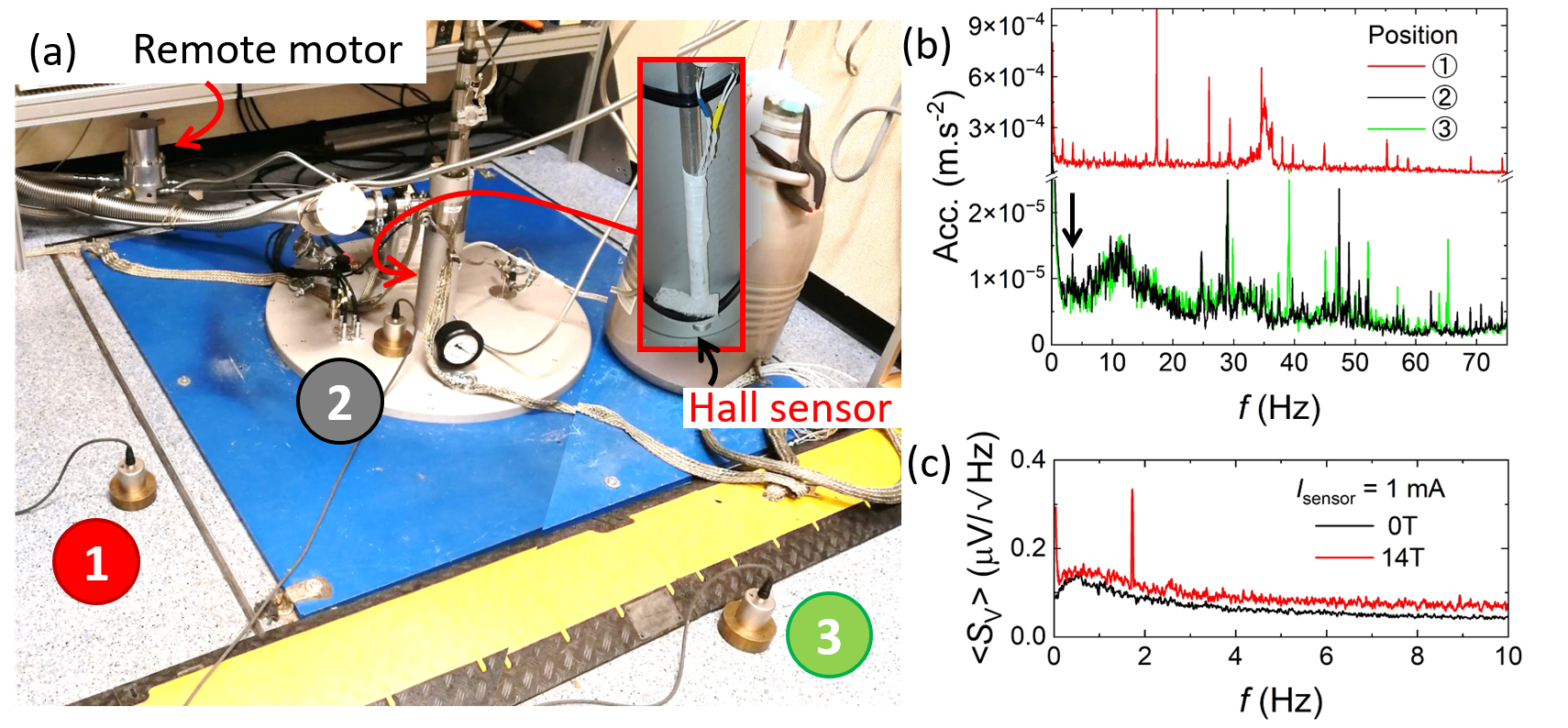}
    \caption{Vibration measurements. (a) Cryomagnetic system and its immediate surrounding, with the position of the accelerometers and the Hall sensor attached to the insertion column. (b) Resulting vibration level at the three different positions. Note the break on the y-axis. The arrow shows the signal at 3.4~Hz. (c) Noise spectral density of the Hall sensor signal with a DC excitation current of 1~mA, at a magnetic field of 14~T and 0~T.}
    \label{fig_vib_acc}
\end{figure*}

The vibration level is estimated using \textcolor{black}{piezoelectric} accelerometers (model PCB 393B31) placed at three different \textcolor{black}{positions} shown in Fig.~\ref{fig_vib_acc}(a): Approximately 1.2~m away from the remote motor on the laboratory floor (position 1), on the top flange of the cryostat (position 2), and approximately 2.3~m away from the remote motor on the laboratory floor, with the cryostat pit in between (position 3). The corresponding results are reported in Fig.~\ref{fig_vib_acc}(b). As expected, the vibration level close to the remote motor (position 1, red curve) is significantly larger than the one on the top flange, in the decoupled pit (position 2, black curve). The vibration level of the laboratory floor slightly away (position 3, green curve) is very similar to the one on the top flange\textcolor{black}{: The decoupled pit is thus not essential to damp the vibrations from the remote motor, and less drastic measures such as moving the cryostat a bit further away from it or placing the remote motor on a dedicated position (e.g. on a support with vibration-dampening features \cite{Chae_2025}) would be enough to reach the background level}.

\textcolor{black}{In the spectrum position 1, a signal at around 1.7~Hz with several harmonics is present}: They correspond to the working frequency of the valve unit (the remote motor) at which the high pressure helium gas is injected/recovered into/from the cryocooler. On the top flange, position 2, a clear signal is observed at 3.4~Hz (its first harmonic) which is absent in the measurement position 3. Therefore, despite \textcolor{black}{the gap between the cryostat valve unit} and the decoupled pit, some vibrations are transmitted to the cryostat, possibly through the cold head \cite{Tomaru_2004, Schmidt_2020, Sato_2025}, and thus are complex to damp further. In order to verify if these ``internal'' vibrations are also transmitted to the superconducting magnet, we \textcolor{black}{set up} a contactless technique using a Hall sensor, tightly attached to the insertion column, as shown in Fig.~\ref{fig_vib_acc}(a). Since the superconducting magnet is only 35~cm below the floor level, the stray field is rather high (around 150~mT at the base of the column for an applied magnetic field of 10~T) and so is the sensor signal. \textcolor{black}{The frequency response of the Hall sensor voltage, using a dynamical signal analyzer (Agilent 35670A), is evaluated at zero applied magnetic field and at 14~T, see Fig.~\ref{fig_vib_acc}(c)}. A clear signal at 1.7~Hz is present at 14~T. This is consistent with report on pulse-tube cryocooler in which the second stage has a peak-to-peak displacement in the order of few tens of micrometers \cite{Tomaru_2004, Schmidt_2020, Sato_2025}, and suggests that the superconducting magnet \textcolor{black}{may be} not perfectly tight to the cryostat frame and consequently the DC magnetic field is inherently modulated at the working frequency of the remote motor. 

\section{The measuring insert}
\subsection{Description}

This cryomagnetic system is used for low-noise high-precision electrical measurements, both in the DC and alternating current (AC) regimes, motivated by future developments of the quantum Hall impedance standard up to few tens of kilohertz \cite{Kalmbach_2014}. For this purpose and to be compatible with the low-temperature high-field conditions, \textcolor{black}{we design} a home-made top-loading insert, using coaxial cables (SM50 from Axon), around 120-cm long with typically 1.5~$\Omega$ for the (copper) inner conductor, 8~$\Omega$ for the (stainless steel) outer conductor and a total capacitance of 160~\textmu F between the inner and outer conductors (connectors and extension included, see below). These cables resistances are \textcolor{black}{small enough to be adapted for precision measurements, and sufficiently large to obtain} small thermal leak between room temperature and the cold finger. The connectors are SMA-type at room temperature and SSMCX-type on the cold finger. The polytetrafluoroethylene (PTFE) dielectric insulation of the cables and connectors ensures a large leak resistance, essential for high-accuracy measurements, and has been verified to be larger than 100~T$\Omega$ in the whole temperature range in the DC regime. The insert can host two TO-8-type sample holders (12 channels each).

\begin{figure}
    \centering
    \includegraphics[width=1\linewidth]{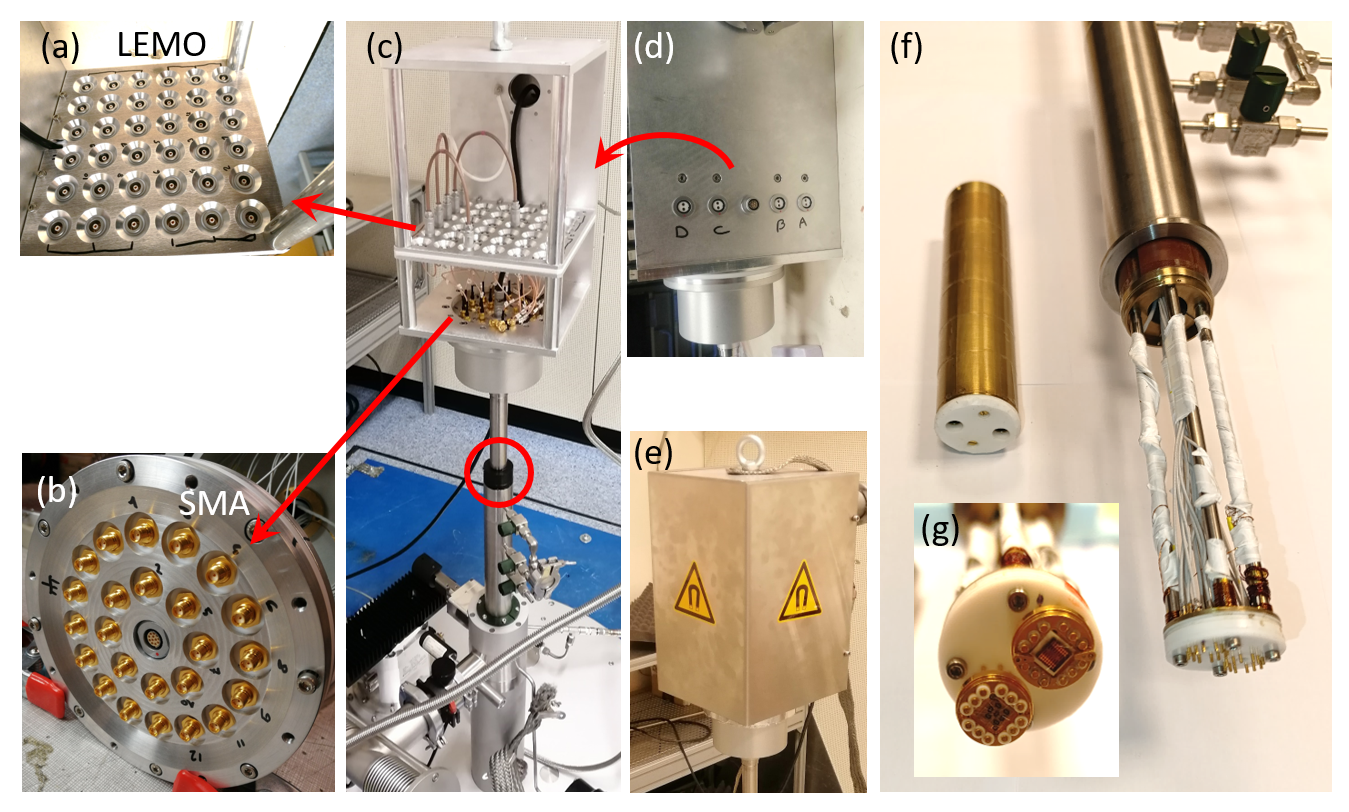}
    \caption{The insert assembled. (a) The LEMO connectors for a practical use, which are solely an extension of the SMA connectors on the head of the insert (b). (c) Insert installed on the cryostat, open to have access to the connectors, or close as shown in (e). The red circle shows the position where the insert lies on the cryostat. (d) Four connectors A, B, C, D placed on the back of the head are used to the split lines for careful grounding, and the connector for the technical lines in the middle. (f) The assembled cold finger with two TO-8 sample holders installed as shown in (g).}
    \label{fig_probe_complete}
\end{figure}

To ensure the coaxiality on the entire insert, the shield of the coaxial cables is shared only on the cold finger on a \textcolor{black}{common} brass plate (decoupled from the rest of the insert with a cotton-reinforced bakelite junction), and on the head the SMA connectors are separated using a PTFE plate (sandwiched between two aluminum plates for mechanical rigidity). Because the SMA connectors lack flexibility in routine operations, a transition from SMA to unipole push-pull PTFE-insulated LEMO connectors is added (also separated by a PTFE plate to keep coaxiality). To minimize the electromagnetic perturbations, the connectors are in an aluminum box which can be closed with a stainless steel hood. Split lines are added to safely ground the devices, and, to limit interferences, the technical lines (temperature and heater leads) are connected in a separated connector. The insert is presented in Fig.~\ref{fig_probe_complete}.

\subsection{Estimation of the vibration level}
\subsubsection{Using Hall sensor}

\begin{figure}
    \centering
    \includegraphics[width=1\linewidth]{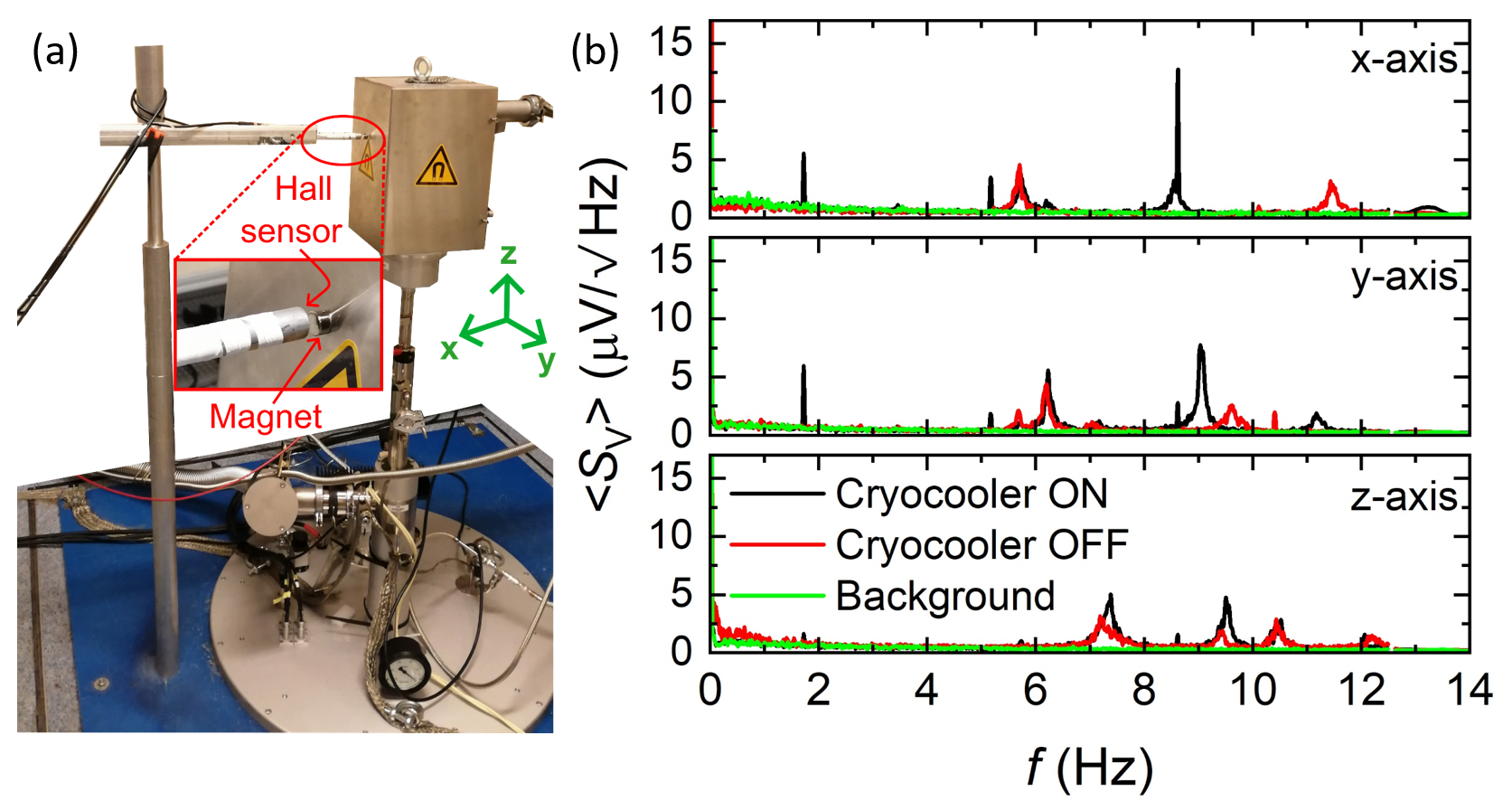}
    \caption{(a) Picture of the setup: A Hall sensor is tied to an arm itself strongly fixed to the steel plate supporting the cryostat. Directly facing the Hall sensor is placed a magnet on the head of the insert. The oscillations have been measured along the three space directions, and when the magnet is away (background). (b) Noise spectral density of the Hall sensor signal along the three directions with the background (in green), when the compressor and remote motor are running (in black) or shut down (in red).}
    \label{fig_vib_Hall}
\end{figure}

\textcolor{black}{The movements of the insert are first evaluated using a Hall sensor and a permanent magnet similarly to the measurement presented above}: The Hall sensor is attached to a pole solidly tied to the steel plate supporting the cryostat, facing a permanent magnet placed on the head of the insert, as shown in Fig.~\ref{fig_vib_Hall}(a). The distance of the Hall sensor to the magnet is \textcolor{black}{adjusted to assure a DC Hall voltage around 20~mV at a current of 1~mA, the noise spectral density of this signal is measured directly by a dynamic signal analyzer}. This operation has been repeated for the three spacial directions, with the cryocooler shut-down (compressor and remote motor off) and under normal operation, and with the permanent magnet away for reference. The results are shown in Fig.~\ref{fig_vib_Hall}(b). When the cryocooler is off (red lines), several peaks are present, corresponding to the ``natural'' oscillations of the setup (insert or pole). When the system is running, additional sharps peaks appears at 1.7~Hz and few harmonics. These signals are mainly along the horizontal directions (x and y), with only a tiny component along the vertical (z) axis. This suggests pendulum-like movement caused by the cryocooler, possibly around the point situated at the top of the insertion column (on the black ring inside the red circle below the head of the insert of Fig.~\ref{fig_probe_complete}(c)) as the insert is not attached to the cryostat but only lies on it. Nevertheless, one can notice the absence of broad signal such as the ones from the vibrations of the top flange (Fig.~\ref{fig_vib_acc}(b)), even at higher frequencies (not shown). This means that, except for these few resonances, the insert \textcolor{black}{have the same movement as} the cryostat, i.e. it does not have a chaotic movement as could have been suspected because the insert is not tightly attached to the cryostat.

\subsubsection{Eulerian Video Magnification}

To go further, we perform a contactless, global acquisition method using high-speed video: the Eulerian Video Magnification (EVM), a computational video processing technique that aims at revealing subtle variations in a video that are otherwise imperceptible to the naked eye \cite{Wadhwa_2016,Luo_2024}. The principle is to monitor the luminance of each pixel of a video, which can vary with time if an object is vibrating. A fast-Fourier transform (FFT) analysis is performed on the pixel luminance to reveal characteristic frequencies (more information can be found in the Supplementary Materials \cite{Taupin_SM}). The recording camera can be placed far away from the elements to be studied (preventing transmitted vibrations), and can easily determine the \textcolor{black}{oscillations} of small objects, such as cables, otherwise extremely difficult, if not impossible, to probe using other methods.
The videos were captured at $f_{acq}=$ 400 frames per second during 30 seconds at a distance of 3 meters using a 4~Mpx Phantom Veo640 camera mounted on a tripod, placed on another decoupled pit similar to the one on which the cryostat lies. This camera, with a 10~\textmu s exposure time, necessitated the use of two 1500~W~LED lights. The video treatment is performed using our own software adapted from \cite{Wadhwa_2014}. The image extraction is done by ffmpeg library, and the computation and video creation in a Labview routine. Two videos were taken: One of the head of the insert (without cover) and one of the top flange of the cryostat \cite{Taupin_2025a}. A screenshot of the video at both positions is shown in Fig.~\ref{fig_vib_video}(a) and in the Supplementary Materials \cite{Taupin_SM}. For any pixel coordinate a frequency spectrum can be extracted (see Fig.~\ref{fig_vib_video}(b) at two selected positions). In these conditions, we can rebuild an image for each frequency, eventually all of these images were stacked versus frequency to rebuild a video \cite{Taupin_2025a}. Fig.~\ref{fig_vib_video}(c) is the reconstructed image in which the color corresponds to the frequency at which the amplitude is maximum in the probed range (1~Hz to 45~Hz), with the panels (d), (e) and (f) the reconstructed images corresponding to the frequencies 1.7~Hz, 7.5~Hz and 40~Hz respectively. Possibly the signals at 7.5~Hz and 40~Hz originate from the ``natural'' vibrations, as signals at similar frequencies have been reported above (see Figs.~\ref{fig_vib_acc}(b) and \ref{fig_vib_Hall}(b)). This experiment however confirms that the insert is oscillating at a frequency of 1.7~Hz, i.e. that vibrations from the remote motor are transmitted to the insert. Importantly, the movement of \textcolor{black}{the insert and of its (loose) cables} is the same, i.e. the cables do not have their own movement (at a higher harmonic for example). Would have it be the case, this could have been a noise source and solutions to tighten the cables would have had to be found.

\begin{figure}
    \centering
    \includegraphics[width=1\linewidth]{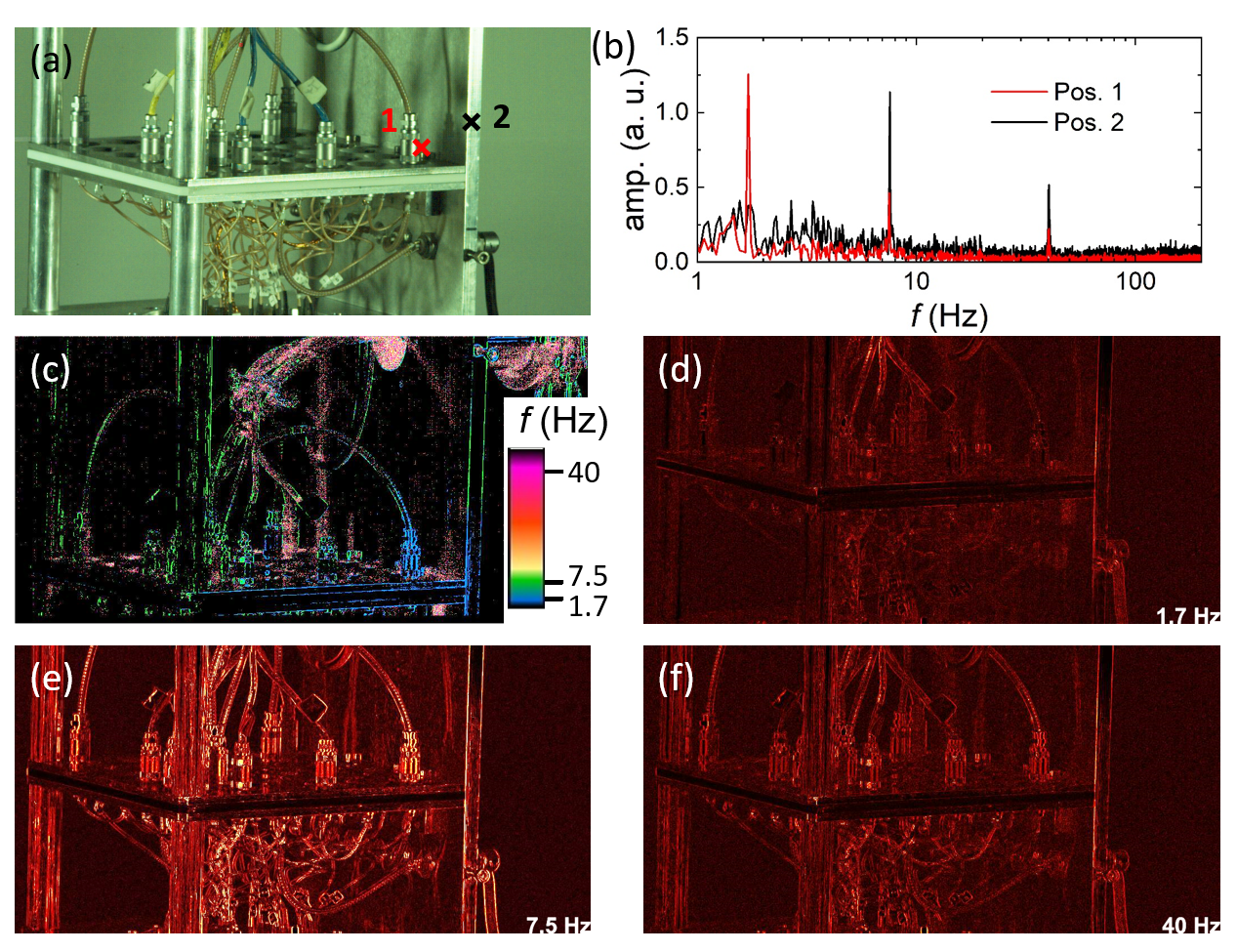}
    \caption{(a) Screenshot of the video taken of the head of the insert. (b) Frequency spectrum of the luminance of the pixel at the positions 1 and 2. (c) Mapping of the vibrations of the entire head. The color corresponds to the frequency at which the amplitude is maximum. The panels (d), (e) and (f) are the mapping of the luminance respectively at 1.7~Hz, 7.5~Hz and 40~Hz.}
    \label{fig_vib_video}
\end{figure}

\subsection{Electromagnetic pollution}

\begin{figure*}
    \centering
    \includegraphics[width=1\linewidth]{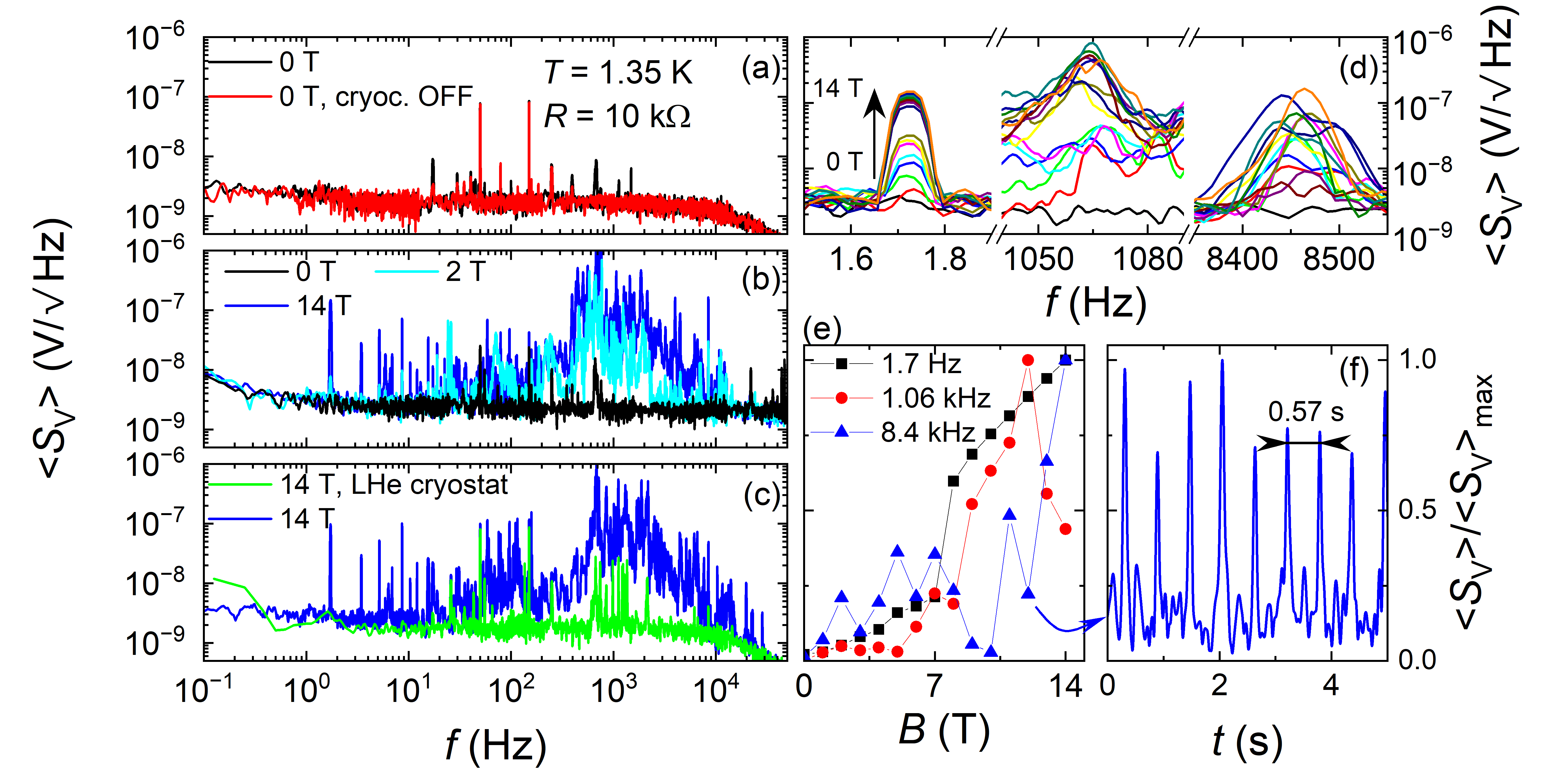}
    \caption{Noise spectral density of a 10~k$\Omega$-resistor at 1.35~K under various conditions. (a) At zero applied magnetic field with the cryocooler turned on (in black) and off (red). (b) Without magnetic field (black) and at 2~T (cyan) and 14~T (blue). (c) At 14~T in our cryogen-free system (blue) and in the liquid-helium-based cryostat (LHe cryostat, green). (d) Evolution of the signal with the magnetic field around selected frequencies, with their normalized amplitude panel (e). (f) Temporal evolution of the signal at around 8.4~kHz at 10~T, normalized by its larger value.}
    \label{fig_10kohm}
\end{figure*}

Vibrations \textcolor{black}{can be source of} electromagnetic perturbations. \textcolor{black}{To evaluate them in our system, we measure} the Johnson noise of a 10~k$\Omega$ resistor (Vishay Z-serie) placed on a TO-8 sample holder at low temperature and under magnetic field. The noise spectral density was measured using the dynamic signal analyzer and a voltage pre-amplifier Celians EPC1-B (with a gain of 100 or 10000). The results are presented in Fig.~\ref{fig_10kohm}. Without applied magnetic field, panel (a), the noise spectrum is rather frequency independent. There are some sharp resonances, notably the 50~Hz from the electrical network and few harmonics. The only differences between the cryocooler being turned on or off are the sharp but moderate-in-amplitude resonances at around 680~Hz and 1~kHz (no signal observable at 1.7~Hz). 
When the magnetic field is applied, at 14~T, the amplitude of the perturbations on the noise spectrum drastically increases \textcolor{black}{and becomes} much larger than the one observed in a liquid-helium-based cryostat under similar conditions (LHe cryostat, panel (c)). A clear resonance at 1.7~Hz is observed with several harmonics, and the perturbations at around 100~Hz and 1~kHz, present in the liquid-helium-based cryostat, are greatly amplified already at low field: At 2~T (panel (b)), the global shape of the noise spectrum is similar to the one at 14~T, only differing in the amplitude of the perturbations. This is \textcolor{black}{evidenced} by the field dependence of the spectra, shown panel (d) around three selected frequencies (1.7~Hz and 1.06~kHz with its likely harmonic at 8.4~kHz): The amplitude of the signals is increasing with the magnetic field (panel (e)). However, while the signal at 1.7~Hz is well-defined and its amplitude increases monotonously with the field, the behavior is more chaotic at 1.06~kHz and 8.4~kHz, in their shape, their characteristic frequency and the field dependence of their amplitude. The reason is likely that the source of these perturbations is itself not well-defined and varies with time, which is verified by using a signal analyzer (Keysight PXA N9030B) to measure the temporal dependence of the amplitude of the signal at around 8.4~kHz: A clear modulation is present, with a period of 0.57~s, i.e. at a frequency of 1.7~Hz, corresponding to the working one of the remote motor. This result strongly suggests that the cryocooler is the direct cause of these electromagnetic perturbations, which are amplified with the magnetic field. Because the amplitude is not constant with time and the peaks are rather short (66~ms on average), the origin of these perturbations seems to have an ``on-off'' behavior. One plausible explanation is that each gas injection into the cryocooler (occurring at 1.7~Hz) create a shockwave in some mechanical parts \cite{Sato_2025}, which translates into electromagnetic perturbations at high frequencies (kilohertz).

\section{Electrical measurements}
\subsection{Magnetoresistance}

\textcolor{black}{In this section we compare electrical measurements from the cryogen-free cryostat with those from the liquid-helium-based cryostat.} We follow the technical guidelines for reliable DC measurements of the quantized Hall resistance \cite{Delahaye_2003}, using  AlGaAs/GaAs (batch LEP 514, see e.g. ref.~\cite{Piquemal_1993}) and graphene on silicon carbide (G/SiC) \cite{Ribeiro-Palau_2015, Couedo_2026} quantum Hall resistance standards (QHRS) for the magnetoresistance, the DC longitudinal resistance and precision measurements.

\begin{figure}
    \centering
    \includegraphics[width=0.66\linewidth]{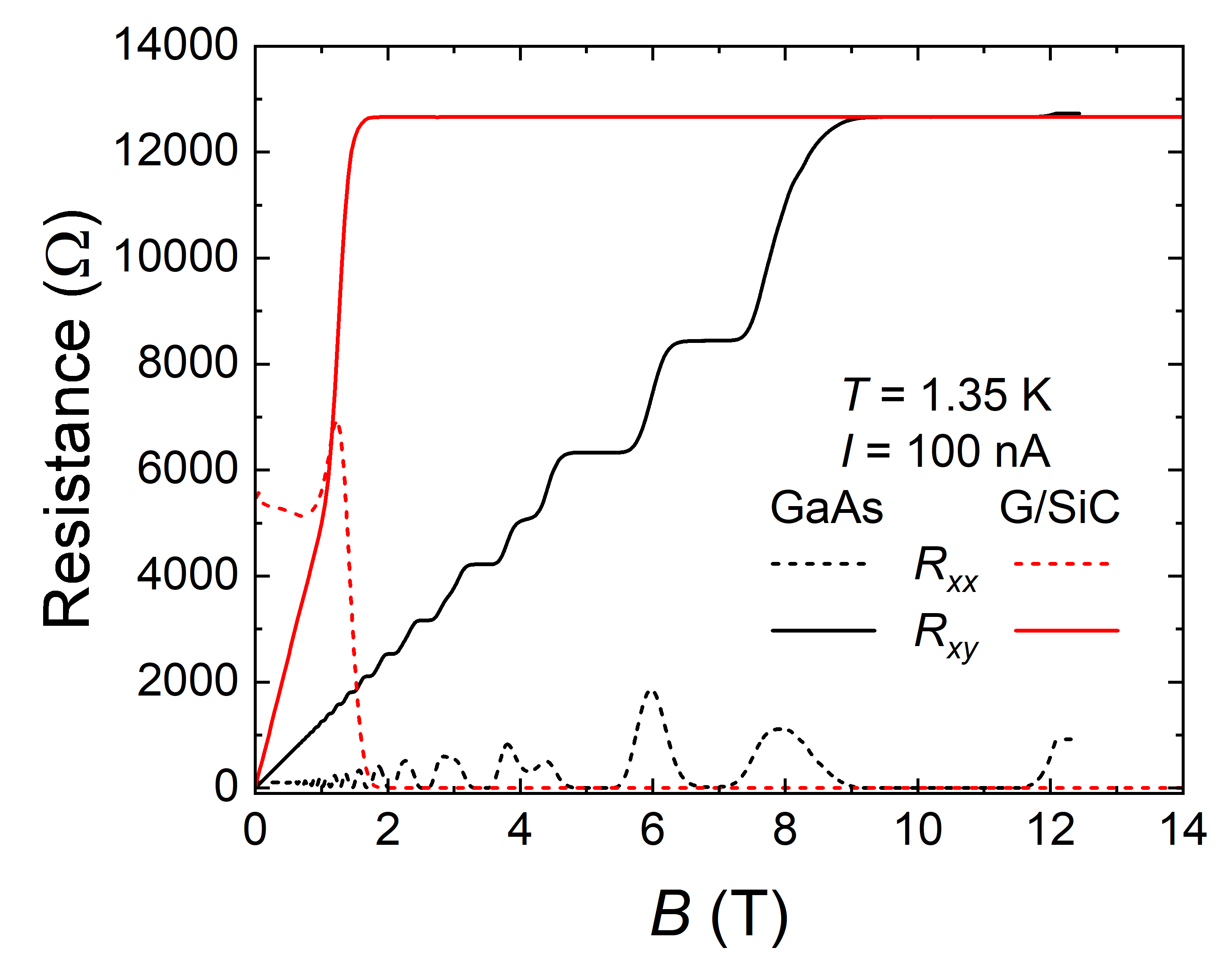}
    \caption{Magnetoresistances at 1.35~K of the AlGaAs/GaAs (in black) and G/SiC (in red) standards. Both the longitudinal (dashed lines) and Hall (continuous lines) resistances are shown.}
    \label{fig_MR}
\end{figure}

The magnetoresistance has been performed in both QHRS at 1.35~K using lock-in amplifiers (Signal Recovery 7265) at 13.33~Hz, Celians EPC1-B preamplifiers and a home-made voltage-to-current converter to apply 100~nA. In the AlGaAs/GaAs device, Fig.~\ref{fig_MR}(a), one observes the typical minima in the longitudinal resistance $R_{xx}$ and the plateaus of the Hall resistance $R_{xy}$  corresponding the different index $\nu$ of the QHE $R_{{H}} = R_{{K}}/\nu$. In the G/SiC device, the plateau at $\nu=2$ is observed from 2~T up to 14~T. \textcolor{black}{We perform measurements} at several fields while staying in the QHE regime to estimate the effect of the magnetic field on the results, all other parameters being kept the same. 

\subsection{DC longitudinal resistance}

\begin{figure}
    \centering
    \includegraphics[width=0.66\linewidth]{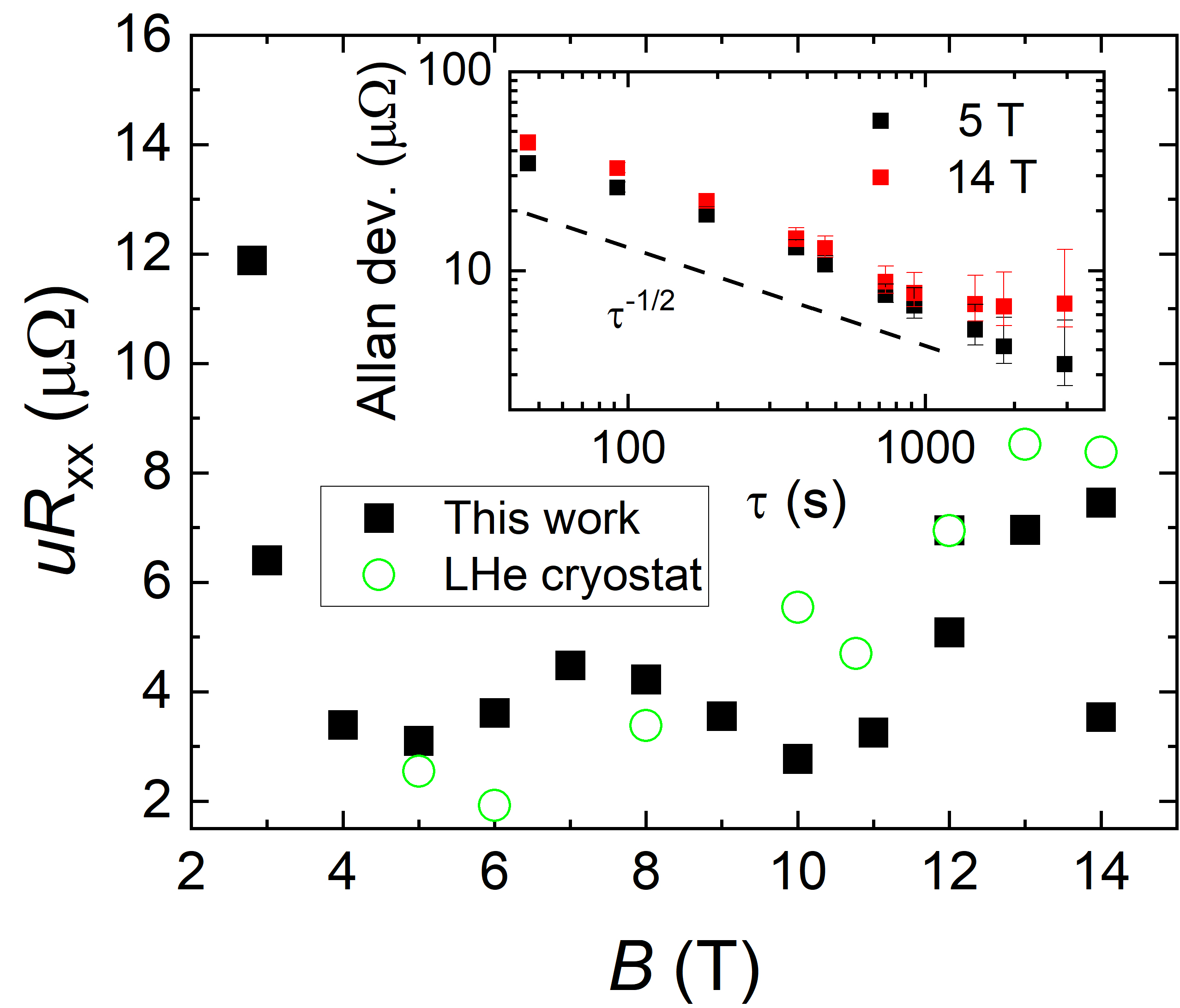}
    \caption{Magnetic field dependence of the measurement uncertainty of the estimate of the longitudinal resistivity (black squares), taken at 1.35~K under a DC current of 50~\textmu A, compared with the measurements taken in the liquid-helium-based cryostat \cite{Ribeiro-Palau_2015} (green open circles). In insert is the Allan deviation taken at 5~T and 14~T. The dashed line illustrates a $\tau^{-1/2}$ dependence, characteristic of white noise.}
    \label{fig_Rxx}
\end{figure}

For the determination of the DC longitudinal resistance at low temperature in the G/SiC device, the current is applied with the home-made current source used for precision measurements \cite{Poirier_2021} and the resulting longitudinal voltage is directly read with a DC nanovoltmeter (N11 from EM Electronics). In the QHE regime, the longitudinal resistance is vanishingly small ($R_{xx}<100~$\textmu$\Omega$) \cite{Couedo_2026}, the interest here lies in the evolution of the measurement uncertainty with the magnetic field. The device is biased under a current of $\pm$~50~\textmu A in quasi-DC conditions (see ref.~\cite{Poirier_2021} for measurement details). The uncertainty $uR_{xx}$, corresponding to the standard deviation (statistical or type A uncertainty), at 1.35~K is shown in Fig.~\ref{fig_Rxx} at various magnetic fields. In the QHE regime ($B > 3$ T), it lies typically between 3~\textmu$\Omega$ and 7~\textmu$\Omega$, \textcolor{black}{similar to the ones obtained using a liquid-helium-based cryostat \cite{Ribeiro-Palau_2015}}, without systematic dependence with the magnetic field. This indicates that despite the increase of the electromagnetic perturbations with the magnetic field observed previously, there is no apparent effect in the noise level of our DC measurements.
We \textcolor{black}{now estimate the overlapped Allan deviation \cite{Allan_1966, Greenhall_1991} of $R_{xx}$ at 5~T and 14~T} using the Stable32 software \cite{stable32}, shown in the insert of Fig.~\ref{fig_Rxx}: Both the amplitude and behavior (proportional to $\tau^{-1/2}$, characteristics of white noise) are similar up to around 1000~s, confirming that the measurement is not significantly affected by the magnetic field. 

\subsection{Precision measurements}
\subsubsection{Validation of the setup}

\begin{figure}
    \centering
    \includegraphics[width=1\linewidth]{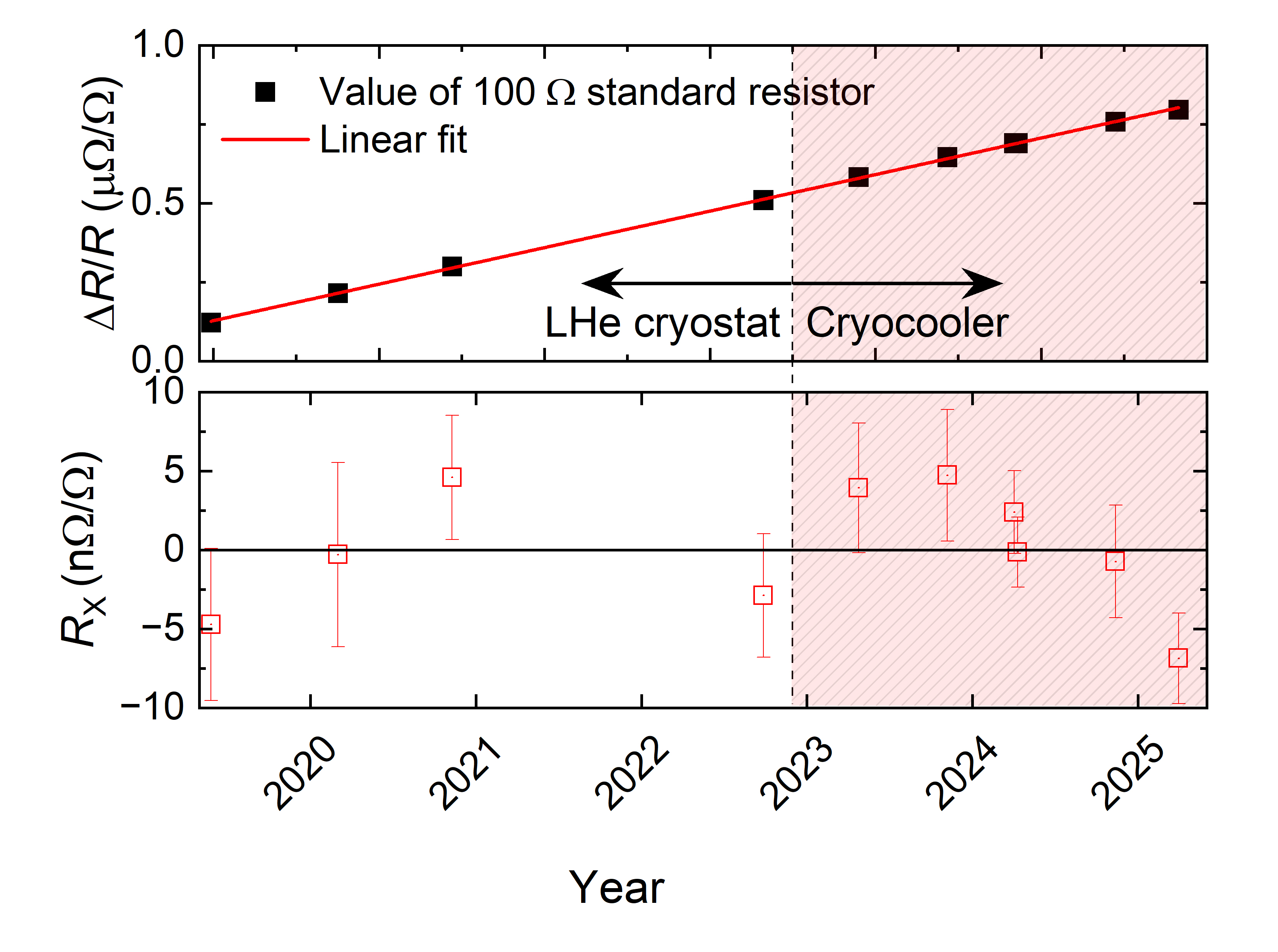}
    \caption{Relative deviation of the 100~$\Omega$ standard resistor (in \textmu$\Omega/\Omega$) compared to a QHRS with the measuring date, depending on the cryomagnetic system used: Before 2023 the liquid-helium-based cryostat (``LHe cryostat'') and after the cryomagnetic system  (``Cryocooler''). The red line is the linear fit of the temporal drift of the resistance value. The bottom panel shows the residual $R_{{X}}$ to a linear fit in n$\Omega/\Omega$. The error bars correspond to the combined uncertainty.}
    \label{fig_suivi_res}
\end{figure}

Next, we perform precision measurements using either the AlGaAs/GaAs or the G/SiC QHRS at 1.35~K, at various magnetic fields in the latter. We compare the value of a well-known 100~$\Omega$ transfer resistance (TEGAM SR102 serie), obtained using the QHRS, depending on the measurement year, \textcolor{black}{using} a home-made resistance bridge associated with its cryogenic current comparator (CCC) \cite{Poirier_2021}, cooled down in a separated cryostat. Before January 2023 was used a liquid-helium-based cryostat (``LHe cryostat'') and after this date this new cryomagnetic system (``Cryocooler''). The value of the 100~$\Omega$ resistance with time and using both cryogenic systems is presented in Fig.~\ref{fig_suivi_res}, \textcolor{black}{with the top panel showing} the relative deviation of the resistance value from its nominal value (in \textmu$\Omega/\Omega$), and the lower one the residual value \textcolor{black}{$R_{{X}}$} from the linear depedence (red line). The residual shows a dispersion within few n$\Omega/\Omega$ without apparent break at the change of cryostat. These results demonstrate that cryogen-free systems deliver state-of-the-art performances \textcolor{black}{for electrical measurements}, similar to the one obtained in an liquid-helium-based cryostat.

\subsubsection{Field dependence of the measurement uncertainty}

\begin{figure}[!ht]
    \centering
    \includegraphics[width=1\linewidth]{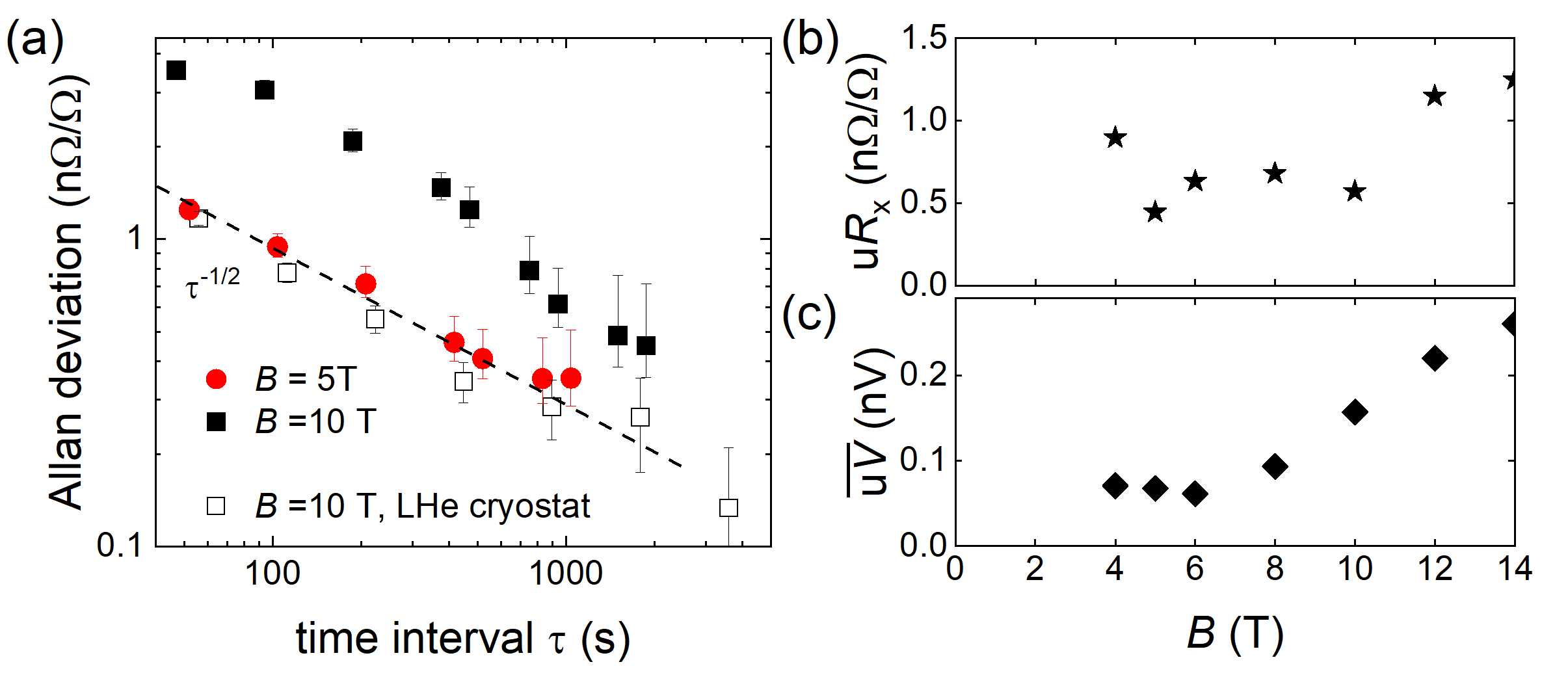}
    \caption{(a) Allan deviation \textcolor{black}{of the precision measurements $R_{{X}}$} in n$\Omega/\Omega$ at 5~T and 10~T together with the one of the liquid-helium-based cryostat at 10~T (open square symbols) \cite{Ribeiro-Palau_2015}. The dashed line illustrates a $\tau^{-1/2}$ dependence. (b) Relative standard deviation $uR_{{X}}$ in n$\Omega/\Omega$ of the value of a 100~$\Omega$ resistance standard with the QHRS at different magnetic fields (measurement duration of approximately 1000~s). (c) Average standard deviation of a single voltage measurement (duration $\sim$ 90~s). For these measurements, a graphene-based QHRS at 1.35~K has been used.}
    \label{fig_typeA}
\end{figure}

\textcolor{black}{We now look into the field dependence of the statistical uncertainty when performing precision measurements $R_{{X}}$. The Allan deviations of $R_{{X}}$ at 5~T and 10~T are shown in Fig.~\ref{fig_typeA}(a) together with the one at 10~T using the liquid-helium-based cryostat \cite{Ribeiro-Palau_2015}: For precision measurements, the Allan deviations are significantly different at 5~T and 14~T highlighting the strong effect of the magnetic field}. It is more pronounced for short measurements (\textcolor{black}{at 100~s, it is 3 times larger at 10~T} than at 5~T) than long ones (1.6 times larger at 1000~s). This is consistent with experimental results: \textcolor{black}{No systematic effect of the magnetic field is observed on the  \textcolor{black}{final} measurement uncertainty ($uR_{{X}}$) as} shown in Fig.~\ref{fig_typeA}(b) (similar uncertainties between 5~T and 10~T \textcolor{black}{for a measurement duration of 1000~s}). \textcolor{black}{For} single data points (time scale of typically 90~s), \textcolor{black}{the magnetic field clearly increases the measurement uncertainty above 6~T}, see Fig.~\ref{fig_typeA}(c). Below, it becomes field independent and the Allan deviation at 5~T matches the one using the liquid-helium-based cryostat (at 10~T), both having a $\tau^{-1/2}$ dependence. Interestingly, this threshold field \textcolor{black}{value} of 6~T is also the one where a discontinuity is observed in the amplitude of the resonance at 1.7~Hz in the noise spectrum density of the 10~k$\Omega$ resistor (Fig.~\ref{fig_10kohm}(e)). This means that above 6~T, the magnetic field (and by extension the electromagnetic perturbations generated by the cryocooler) becomes the dominant noise source in these measurements.

\subsubsection{Effect on the DC-SQUID}

\begin{figure}
    \centering
    \includegraphics[width=1\linewidth]{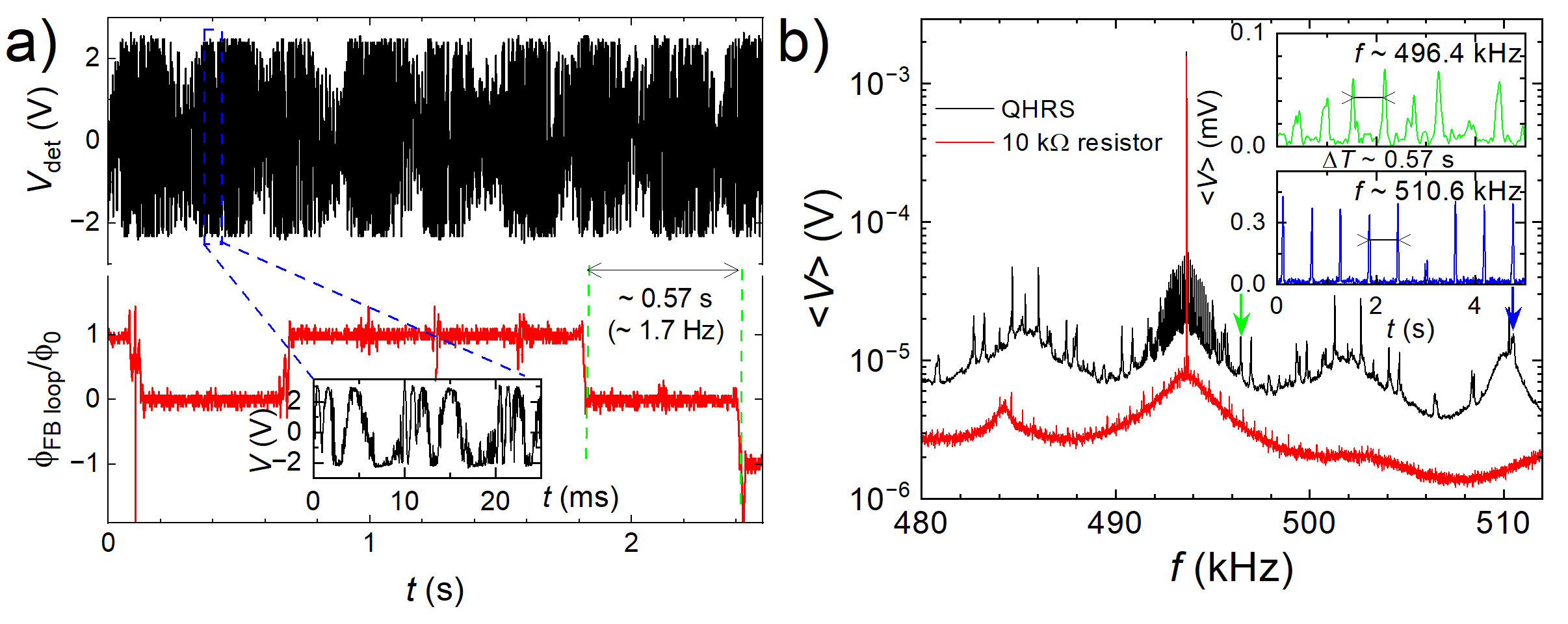}
    \caption{SQUID response with the QHRS at 1.35~K and 10~T. (a) On top, the signal waveform from the detector over a 2.5~s duration (in inset is a detail on 25~ms) and on the bottom the measured flux $\phi$ divided by the quantum flux $\phi_0$ (the position $\phi=0$ is arbitrary). (b) Frequency dependence of the SQUID detector signal at the vicinity of the SQUID modulation, using the QHRS (in black) and a 10~k$\Omega$-resistor (in red). The temporal evolution of the amplitude of the signal at around 496.4~kHz and 510.6~kHz (green and blue arrow respectively) is shown in inserts.}
    \label{fig_spectre_SQUID}
\end{figure}

To pinpoint the reason why the magnetic field affects these measurements, \textcolor{black}{we look at the CCC, and more precisely its DC-SQUID used} to stabilize the current of our resistance bridge for the precision measurements (see ref.~\cite{Poirier_2021} for details). Together with the increased uncertainty, we also noticed a reduction of the stability of the DC-SQUID \textcolor{black}{as the  magnetic field is increased}, which manifests itself as an abrupt change of the working point by a quantum flux. The effect of the cryocooler can be seen in the waveform from the detector and the measured flux of the SQUID controller (model 550 from Quantum Design) as shown in Fig.~\ref{fig_spectre_SQUID}(a): With the QHRS at 1.35~K and 10~T, a clear modulation is observed on the signal of the SQUID detector, as well as periodic ``shakings'' of the feedback loop signal, which can trigger a flux jump by a quantum flux $\phi_0=h/2e$. These anomalies have a period of around 0.57~s, i.e. a frequency of 1.7~Hz, confirming that they can be linked to the cryocooler.
To understand how the perturbations generated by the cryocooler can effect the DC-SQUID response, we use the signal analyzer to look at the frequency dependence of the SQUID detector at the vicinity of the SQUID modulation frequency (around 500~kHz). The signal is shown in Fig.~\ref{fig_spectre_SQUID}(b) (in black) together with a reference measurement using a 10~k$\Omega$ standard resistor (in red): A clear sharp resonance is present at around 494~kHz (SQUID modulation), but with the QHRS, the signal is polluted with various pockets of anomalies. The amplitude of these perturbations is modulated with time, as shown in the inserts for the peaks at around 496.4~kHz and 510.6~kHz (green and blue arrows respectively): We find again a period of the modulation to be around 0.57~s. The explanation suggested above still hold: Each ``shock'' produced by gas injection in the cold head create a shockwave which generate electromagnetic perturbations also at very high frequency (hundreds of kilohertz). Because these are at the vicinity of the SQUID modulation frequency, interferences could happen, which become more pronounced with the increase of the magnetic field as the overall electromagnetic noise increases (Fig.~\ref{fig_10kohm}). This issue is specific to the DC-SQUID, as no effect of the magnetic field is observed using the nanovoltmeter \textcolor{black}{(previous section)} or using another CCC equipped with a RF-SQUID (modulation frequency of 20~MHz). To prevent these interferences from spoiling our measurements, one needs to use a damping circuit \cite{Poirier_2021}, or to reduce the sensitivity of the SQUID feedback loop. The investigation of the causal links between the cryocooler and the DC-SQUID response goes beyond the scope of the present study.

\section{Conclusion}
A cryogen-free pulse-tube based cryomagnetic system and its home-made coaxial insert have been investigated in depth. The system presents high thermal performance in terms of cooling power and thermal stability. We demonstrate the feasibility of high-precision low-noise electrical measurements including stability with a SQUID feedback loop, even at high magnetic field (up to 14~T), with similar performances compared to liquid-helium-based cryostats. A detailed study of the vibration level has been performed using different technics, implementable in any laboratory (no need of expensive apparatus), some giving the possibility to inspect positions otherwise impossible to do using standard technics (not accessible or too small/soft). Despite several layers of protection to damp the mechanical vibrations of the compressor and the remote motor (compressor in another room, remote motor separated from the cryostat itself in a decoupled pit), the measurement insert shows displacements at the characteristic frequency of the remote motor. These vibrations induce electromagnetic perturbations up to surprisingly high frequency (at least hundreds of kilohertz). Electrical measurements are not necessarily affected by the \textcolor{black}{noise generated by the} cryocooler, but when it is the case, it could be caused by interferences when the instrumentation has a working frequency nearby one of a generated perturbation. As a consequence, to make the most of cryogen-free cryostat, a good knowledge of its perturbations, vibrational and electromagnetic, and of the instrumentation used is essential. 

\begin{acknowledgments}
We thank W. Poirier and F. Schopfer for their initial work on the cryostat and useful discussions, O. Thévenot and A. Imanaliev for help in designing the coaxial insert, M. Mghalfi for technical assistance, and F. Piquemal for enlightment and proof-reading.
This work has been partially supported by the European Union’s Horizon Europe research and innovation programme through the Qu-Test project (HORIZON-CL4-2022-QUANTUM-05-SGA, under Grant Agreement No 101113901), the Joint Research Project 18SIB07 GIQS—Graphene Impedance Quantum Standard, and the EIC Pathfinder project ``FLATS'' No 101099139.
\end{acknowledgments}

\section*{Data Availability Statement}
The data that support the findings of this study are available from the corresponding author upon reasonable request.

\section*{Conflict of interest Statement}
The authors have no conflicts to disclose.

\section*{Supplementary Materials}
Technical details concerning the the Eulerian Video Magnification (EVM) method are available in the Supplementary Materials, together with the video and its analysis of the top flange of the cryostat.

\section*{Author contributions}
\textbf{Mathieu Taupin}: Conceptualization (equal); Methodology (equal); Writing – original draft (lead); Formal analysis (lead); Writing – review and editing (equal).
\textbf{Kamel Dougdag}: Resources (lead); Writing – review and editing (equal).
\textbf{Djamel Ziane}: Software (lead); Writing – review and editing (equal).
\textbf{Fran\c cois Cou\"{e}do}: Conceptualization (equal); Methodology (equal); Writing – original draft (supporting); Formal analysis (supporting); Writing – review and editing (equal).

\providecommand{\noopsort}[1]{}\providecommand{\singleletter}[1]{#1}%

\end{document}